\documentclass{article}
\usepackage{url}

\begin{document}
\title{Edsger Wybe Dijkstra (1930 -- 2002): \\  A Portrait of a Genius}
\author{Krzysztof R. Apt,     \\    
        {\em CWI }, 
        {\em P.O. Box 94079, 1090 GB Amsterdam, The Netherlands} \\
        and \\
        {\em University of Amsterdam, The Netherlands}
\footnote{Currently on leave at the School of Computing, National University of Singapore} 
}

\maketitle

\section*{Scientific Career}

Edsger Wybe Dijkstra was born in Rotterdam May 11, 1930.  His mother
was a mathematician and father a chemist. In 1956 he graduated from
the University of Leiden in mathematics and theoretical physics. In
1959 he received his PhD from the University of Amsterdam for his
thesis titled ``Communication with an Automatic Computer'' devoted to
a description of the assembly language designed for the first
commercial computer developed in the Netherlands, the X1.  It also
dealt with the concept of an interrrupt, a novelty at that time.  His PhD
thesis supervisor was Aad van Wijngaarden.

From 1952 till 1962 he worked at the Mathematisch Centrum in Amsterdam
where he met his wife Ria. In 1962 they moved to Eindhoven where he
became professor at the Mathematics Department at the Technical
University of Eindhoven.  Then in 1964 they moved to a newly built
house in Nuenen, a small village on the outskirts of Eindhoven which
in 1973 was added to the world map of computer science when Dijkstra
started to circulate his reports signed ``Burroughs Research Fellow''
with his home address.  Many thought that Burroughs, a company known
at that time for the production of computers based on an innovative
hardware architecture, was based in Nuenen.  In fact, Dijkstra was the only
research fellow of Burroughs Corporation and worked for it from home,
occasionally traveling to its branches in the USA.

As a result he reduced his appointment at the University to one day a
week.  This day, Tuesday, soon became known as the day of the famous
``Tuesday Afternoon Club'', a seminar during which he discussed with
his colleagues scientific articles, looking at all aspects --
notation, organization, presentation, language, content, etc. 
Soon after he moved in 1984 to the University of Austin, Texas, USA, 
a new ``branch'' of the Tuesday Afternoon Club emerged in Austin.
Dijkstra worked in Austin till his retirement in the fall of 1999.  He
returned from Austin, terminally ill, to his original house in Nuenen
in February 2002, where he died half a year later, on August 6th. He
is survived by his wife and three children, Marcus, Femke and Rutger. 

\section*{Scientific Contributions}

Through his fundamental contributions Dijkstra shaped and influenced
the field of computer science like no other scientist. His 
groundbreaking contributions ranged from the engineering side of computer
science to the theoretical one and covered several areas including
compiler construction, operating systems, distributed systems,
sequential and concurrent programming, software engineering, and graph
algorithms.  Many of his papers, often just a few pages long, are
the source of whole new research areas.  Even more, several concepts
that are now completely standard in computer science were first
identified by Dijkstra and bear names coined by him.

Examples abound.  In 1959 he published in a 3-page long article ``A
Note on Two Problems in Connexion with Graphs'' the celebrated,
supremely simple, algorithm to find the shortest path in a graph, now
called Dijkstra's algorithm.  Its impact over the next forty years is
best summarized by the following quotation from the article of Mikkel
Thorup, ``Undirected Single-Source Shortest Paths with Positive Integer
Weights in Linear Time'', from Journal of the ACM 46(3), pp.362-394, 1999:

\begin{quote}
``Since 1959, all theoretical developments in SSSP [Single-Source
Shortest Paths] for general directed and undirected graphs have been
based on Dijkstra's algorithm.''
\end{quote}

Following Fortran, ALGOL 60 was the second high-level programming
language. Dijkstra was closely involved in the ALGOL 60 development,
realization and popularization. As discussed by Peter Naur in the
article ``The European side of the last phase of the development of
ALGOL 60'', in the Proceedings of the first ACM SIGPLAN conference on
History of programming languages, January 1978, Dijkstra took part in
the period 1958--1959 in a number of meetings that culminated in the
publication of the report defining the ALGOL 60 language.  Dijkstra's
name does not appear in the list of thirteen authors of the final
report.  Apparently, he eventually left the committee because he could
not agree with the majority opinions.  Still, while at the
Mathematisch Centrum, he wrote jointly with Jaap Zonneveld the first
ALGOL 60 compiler.  It employed a novel method for implementing
recursion.  His short book ``Primer of Algol 60 Programming'',
originally published in 1962, was for several years the standard
reference for the language.

In a one page paper from 1965 he introduced the ``mutual exclusion
problem'' for $n$ processes and discussed a solution to it.  It was
probably the first published concurrent algorithm.  The notion,
standard by now, of a ``critical section'' was also introduced in this
paper.  The 1986 book by Michel Raynal, titled ``Algorithms for Mutual
Exclusion'', shows what impact this single page had on the field in
the first twenty years since it was published.

Dijkstra and his colleagues in Eindhoven also designed and implemented
THE (standing for ``Technische Hogeschool Eindhoven'') operating
system, which was organized into clearly identified layers.  His 1968
article on this subject provided the foundation for all subsequent designs of
the operating systems.

In 1968 Dijkstra published his famous paper ``Cooperating Sequential
Processes'', a 70-page essay that originated the field of
concurrent programming.  He discussed in it the notion of
mutual exclusion and the criteria a satisfactory solution should
satisfy. He also redressed the historical perspective left out of his
1965 paper by including the first solution to the mutual exclusion
problem, for 2 processes, due to Th.J. Dekker.  Further, he proposed
the first synchronization mechanism for concurrent processes, the
semaphore with its two operations, P and V.
 He also identified the ``deadlock problem'' (called there
``the problem of the deadly embrace'') and proposed an elegant
``Banker's algorithm'' that prevents deadlock.  The deadlock detection
and prevention became perennial research problems in the field of
concurrent programming.

Several of these ideas were conceived by him much earlier. For
example, he introduced the concept of a semaphore already in 1962.
It is discussed in his manuscript EWD 51, ``Multiprogrammering en
X8'' (Multiprogramming and X8) written in Dutch, see
\url{http://www.cs.utexas.edu/users/EWD/ewd00xx/EWD51.PDF}.  The P and
V operations are abbreviations for ``Prolaag'', a non-existing Dutch
word that stands for ``lower'' (in Dutch ``lower'' is ``verlaag'') and
for ``Verhogen'', a Dutch word for ``raise''.  In turn, the ``Banker's
algorithm'' appeared in his manuscript EWD 108, ``Een algorithme ter
voorkoming van de dodelijke omarming'' (An algorithm to prevent the
deadly embrace) written in Dutch, see
\url{http://www.cs.utexas.edu/users/EWD/ewd01xx/EWD108.PDF}.  Also,
the paper ``Cooperating Sequential Processes'' was finalized in 1965
and was available as his manuscript EWD 123, see
\url{http://www.cs.utexas.edu/users/EWD/ewd01xx/EWD123.PDF}.

Then, in a 1971 paper he illustrated the deadlock problem by means of
the ``dining philosophers problem'' according to which five
philosophers, seated around a table, are supposed to eat spaghetti
sharing only five forks.  This problem became a classic benchmark for
explaining new synchronization primitives.  The paper also led to an
intense research for high-level synchronization mechanisms, leading
eventually to the concept of a monitor, due to Per Brinch Hansen and
Tony Hoare.

His two-page article ``Self-stabilizing systems in spite of
distributed control'' from 1974 is at the source of one of the main
approaches to fault tolerant computing, as can be seen by studying the
book ``Self-stabilization'' (the name was coined by Dijkstra) of Shlomi
Dolev from 2000 and by browsing through the proceedings of the annual
workshops on self-stabilizing systems. Interestingly, the paper was
noticed only in 1983, after Leslie Lamport stressed its importance in
his invited talk at the ACM Symposium on Principles of Distributed
Computing (PODC). It won the 2002  PODC significant paper award.

Dijkstra had the audacity to criticize the customary ``if-then-else''
programming statement as asymmetric and proposed in 1975 instead
another, symmetric, construct based on the notion of a ``guard''.
This allowed him to present the more than 2300 years old Euclid's
algorithm for computing the greatest common divisor of two natural
numbers in an aesthetically pleasing symmetric form.  Since then the
concept of a guard spread deeply inside computer science.

In 1976 he published a seminal book ``A Discipline of Programming'' which
put forward his method of systematic development of programs together
with their correctness proofs.  In his exposition he used his tiny
``guarded commands'' language. The language, with its reliance on
nondeterminism, the adopted weakest pre-condition semantics, and the
proposed development method has had a huge impact on the field till
this day. 

In 1984, to add further support to this approach to programming, he
published jointly with Wim Feijen an introductory textbook for first
year students of computer science.  The book, first published in
Dutch, was titled ``Een methode van programmeren''. The English
edition appeared in 1988 as ``A Method of Programming''.

In the early eighties he published two small papers on the problem of
detecting termination in distributed systems.  The first one, 4 pages
long, was written with Carel Scholten, the second, 3 pages long, with
Wim Feijen and Netty van Gasteren.  The problem was independently
formulated and solved by Nissim Francez in a more restricted setting
of CSP programs. It became one of the most often studied problems in
the area of distributed programming and a number of surveys on the
subject appeared.

Then in 1990 he published his book ``Predicate Calculus and Program
Semantics'' with Carel Scholten. The book was devoted to logical and
mathematical analysis of his weakest precondition semantics with a
long prelude concerning predicate calculus. However, the book received
mixed reviews. We shall return to this matter later.

His fundamental achievements were early recognized. Already in 1972 he
obtained the ACM Turing award. He was a Foreign Honorary Member of the
American Academy of Arts and Sciences, member of the Royal Netherlands
Academy of Arts and Sciences (KNAW), and held the doctor honoris causa
titles from the Universities of Belfast, Northern Ireland and Athens,
Greece.  In addition, over the period of thirty years, he received
numerous awards and distinctions, some just a couple of weeks before
his death.  On April 10, 2001, the Dutch television broadcast a half
an hour TV program about Dijkstra, which was very favourably reviewed
in the main daily, the NRC Handelsblad.

Not surprisingly, his obituary appeared in a number of newspapers, including
the New York Times, the Washington Post, and Guardian.

\section*{Working Style}

Dijkstra never wrote his articles using a computer. He preferred to
rely on his typewriter and later on his Mont Blanc fountain pen. These
articles were then distributed in an old fashioned way: he sent copies
to a few friends and associates who then served as the source nodes of
the distribution centers. These short articles span over the period
of forty years. They are rarely longer than 15 pages and are
consecutively numbered.  The last one, No 1318 is from April 14, 2002.
Within computer science they are known as the EWD-reports, or,
simply the EWD's.
The early ones are not dated so it is difficult to ascertain their publication date.

A major change occurred when as part of the celebration of his 70th
birthday, the computer science department in Austin issued in 2000 a
CD-rom with most of the EWD-reports. They are also available from the
website \url{http://www.cs.utexas.edu/users/EWD/} maintained by Ham
Richards, with the more recent reports added.  This huge amount of
material, in total over 7700 pages, consists of scientific articles,
essays, position papers, conference and scientific trip reports, open
letters, speeches, and lately, increasingly often, elegant expositions
of solutions to well-known and less well-known combinatorial problems
and puzzles.

Dijkstra applied to his work a rigorous self-assessment procedure and
only a small fraction of the EWD's were eventually submitted to
refereed journals. As a result many of his contributions are not
well-known.

His handwriting was so perfect and distinct that in the late eighties
Luca Cardelli, then from the DEC Systems Research Center, designed a
``Dijkstra'' font for Macintosh computers.  
Soon after, Dijkstra got a letter typeset in this font and thought
it was handwritten until news reached him about the
creation of this font.  Some of Dijkstra's colleagues occasionally
used this font in their slide presentations during the departmental
meetings in Austin.  Those curious to see his striking handwriting and
his extraordinarily elegant exposition style can download for example
his presentation of the perennial wolf, goat and cabbage puzzle (see
EWD 1255 from 2000).

In writing his elegance was unmatched. He could write about formal
issues in the form of an essay, with hardly any formulas.  His,
already discussed, paper ``Cooperating Sequential Processes'' is
perhaps the best example. Similarly, he was able to discuss (one
should rather say, derive) intricate algorithms in distributed
computing in a seemingly informal way, in plain prose, with just a few
simple formulas.  He wrote his articles in a unique style
characterized by conciseness, economy of argument, and clarity of
exposition. Each sentence was carefully chiseled. Each paragraph was
striking.

In fact, in all that he did he was a perfectionist to the utmost.  His
lectures were always impeccably delivered, often with a sense of
unique drama, and given only with a chalk and a blackboard, completely
out of his head. They were also highly entertaining because of his
sharp comments, striking turns of phrase, or curious quotations that
he used to put on the blackboard before starting his lecture.
In a classroom he would never ask an audience to keep silent.
Instead, he would lower his voice to the point of being hardly audible.
This trick was amazingly effective.

Starting in the late seventies he got interested in the subject of the
development and presentation of proofs. Some of these proofs were
surprising applications of his programming methodology to geometry or
algebra.  He criticized the use of implication, an unneeded reasoning
by cases, or the reasoning by contradiction.  Instead, he favoured the
proofs presented as chains of equivalences with each step justified as
an interlaced comment, and liked to stress the fact that the
equivalence is associative, a fact that logicians knew but apparently
never used.

He also designed his own notation for first-order logic that took
better care of the quantifiers with explicitly given ranges and 
in general thought nothing of mathematicians' disdain for presentation
and lack of attention to the notation. He even criticized the familiar
use of the $\Sigma$ for summation as sloppy and
misleading.  Moreover, he repeatedly argued for numbering starting from
zero, so his reports, from a certain moment on, invariably begin with
page 0, and when he was writing about $n$ processes, they were
invariably numbered $0, \dots, n-1$.  Also, he opposed using drawings
or examples to illustrate concepts, for example specific type of
graphs.

\section*{His Opinions}

In personal contacts with colleagues, he tended to appear stiff,
austere and aloof.  On some rare occasions he was even plainly rude,
like in the early eighties when in Utrecht he demonstratively left in
the middle of a lecture on computer networks given by a prominent
computer scientist from the Free University of Amsterdam, but only
after having bombarded him with questions about terminology. Several
years later, in Austin, he said ``Thank God'' in reaction to a comment
``I am losing my voice'' uttered by a renowned computer scientist from
the MIT toward the end of her lecture. (A letter with apologies from
the department chairman promptly followed.)  But in informal, private
meetings in his office, he could be most charming, serving coffee to
students and making subtle one liner jokes of his own creation.

As a result of his reserved behaviour and uncompromising positions
he was disliked by a number of colleagues, in particular in the
Netherlands. They saw nothing in his prophet-like statements and
biting comments, often delivered from the back of the lecture room,
and directed against the hacker's approach to programming, against the
sloppily chosen notation, or expressing his disapproval of the badly
organized lecture.  Often this scepticism towards his opinions and
ideas could be simply explained by plain jealousy that his viewpoints
and articles were widely cited and discussed.  Consequently, when a
prominent speaker was sought for some important event in the
Netherlands, Dijkstra was usually passed over.

On a number of occasions, his extreme views became the standard
currency several years later.  For example, in 1968 he published a
two-page note ``Goto considered harmful'' that criticized the use of
the \texttt{goto} statement in programming languages.  It led to a
huge uproar. Thirty years later the \texttt{goto} statement shines by
its absence in Java, and the title ``idea x or construct y considered
harmful'' has been borrowed several times, most recently in an article
in the Communications of the ACM Vol. 45 (8), 2002.

His opinions on programming were often sharply criticized and hotly
debated, but they paved the way for the increased attention to
programming methodologies and use of formal methods for verifying
software for ``critical applications'', and contributed early to a
better understanding of the complexity involved in the programming
process.  In the end, if a spokesman was sought to elucidate the
problems of the ``software crisis'', the eyes turned invariably
towards Dijkstra.

Dijkstra enriched our vocabulary by numerous other terms, phrases and
slogans.  The widely used term ``structured programming'' was coined
in his elegant ``Notes on Structured Programming'' from 1972.  Another
slogan, ``separation of concerns'', so often used in software
engineering, goes back to his short note ``On the role of scientific thought''
from 1974 (EWD 447).

He could formulate his opinions with utmost clarity and with a
razor-sharp precision, so they lent themselves naturally to be used
as mottos to book chapters or articles, or as a justification for a
new line of research.  One could easily publish a booklet with his
aphorisms and strikingly refreshing statements about computers,
software, or computer science. 

These opinions could also lead to strong opposition from other
computer scientists, including prominent ones. Don Knuth wrote in 1974
a 40 pages long article titled ``Structured Programming with goto
Statements''.  Dijkstra's views on teaching computer science, 
presented during the ACM Computer Science Conference in February 1989
in a talk titled ``On the Cruelty of Really Teaching Computer Science''
led to a publication in the Communications of the ACM Vol. 32 (12), 1989 of ``A
debate on teaching computing science''. In this issue Richard Karp, Richard
Hamming and other prominent computer scientists criticized his
opinions as too extreme and too radical, in particular because of his
insistence that an introductory course in programming should be
primarily a course in formal mathematics, completely free of program
testing.

It should be also added that on some occasions Dijkstra's lack of
reception of new ideas seemed to stem from his inability to see
through a notation he disliked, his rejection of correctness proofs
based on operational reasoning (by this yardstick temporal logic was
``out''), or from his resistance to absorb ideas that were presented
in a highly technical form.  Also, on a couple of occasions, his
refusal to acquaint himself with the basic literature in mathematical
logic led him astray, like his ``discovery'' that equivalence is
associative.  

In fact, his and Scholten's book ``Predicate Calculus and Program
Semantics'' received a devastating review by Egon B\"{o}rger in
Science of Computer Programming 23 (1994), pp. 1-11.  B\"{o}rger
showed how several laws of propositional calculus, the proofs of which
are ``spiced with pompous methodological comments'', can be proved in
a completely straightforward way using the approach of I.I.  Shegalkin
from 1928, in which each Boolean expression is represented as a
polynomial with the coefficients 0 or 1.  B\"{o}rger also harshly
criticized the highly biased account of the development of predicate
logic in which only selected few logicians were mentioned and in which
a straight line was drawn from Leibniz, through Boole to the authors.
To be fair, one should mention here that Dijkstra's and Scholten's
approach to proof presentation, together with a work of others, also
led to a development of what is now known as ``calculational logic''.

Such a cavalier approach to references was one of the reasons why some
colleagues resented Dijkstra.  In fact, bibliographic references were
never a strong point of Dijkstra's work, and most of his articles and
books have no references at all.  In the preface of his book ``A
Discipline of Programming'' he simply stated disarmingly ``For the
absence of a bibliography I offer neither explanation nor apology''.

All that was probably the small price he had to pay for being able to
focus on his own ideas and his own approach.  After all, in most of
what he did he was a pioneer and consequently an autodidact.
Apparently, to remain creative and highly original he had to shut off
other people's work. He would rather derive all results from first
principles, ignoring work done by others.  He was more interested in
the thought process behind the development of a result, rather than
the result itself.  In fact, most of his work has dealt with
methodology.

His strong personality combined with remarkable working habits and
definite opinions on how to conduct research appealed to many
researchers.  He seemed to believe that everyone should think, and
even behave, the way he did. This made him a natural prophet and
accounted for many of his idiosyncrasies.  He attracted a relatively
small but stable group of disciples, which included both PhD students
and highly renowned computer scientists, who adopted his writing style
and notation, his manners, use of a fountain pen, and occasionally
even his type of sandals.  

\section*{Life in Austin}

In Austin he found his second home.  He liked the United States and
its national parks and often spent vacation traveling around with his
wife in their Volkswagen bus, dubbed the Turing Machine.  When
operating on familiar territory, he was sociable and friendly.  He and
his wife would often drop in on friends and colleagues in the evening,
unannounced, for a social half hour.  He was also often most helpful
and with an original sense of humour.  When asked once how many PhD
students he had, he replied with a smile: ``Two.  Einstein had none''.
(Eventually, Dijkstra had four PhD students: Nico Haberman, who for
several years was the head of the Computer Science Department at
Carnegie Mellon University, Martin Rem, who became the president of
the Technical University of Eindhoven, Netty van Gasteren, who till
her recent untimely death worked at the same University, and David
Naumann who works at the Stevens Institute of Technology, in Hoboken,
NJ and who had Dijkstra and Tony Hoare as PhD supervisors.)

He was liked and respected by his colleagues, who were struck by his
unassuming behaviour.  J Strother Moore once remarked about
Dijkstra's collegial attitude during departmental meetings: ``Edsger
is a great faculty member. He believes in the principle `one person,
one vote' and sticks to it''. In fact, Dijkstra never mingled in 
university politics and always stayed outside of conflicts. At the
same time he was extremely perceptive about people and could
immediately recognize who was a dedicated scientist and who was a
disguised politician.

He often attended lectures of invited speakers and would do his
best to closely follow them till the very end.  His courses for
students in Austin had little to do with computer science: they dealt
with the presentation of mathematical proofs.  Actually, on the
Department's home page, one could read the following terse summary of
his research: `` My area of interest focuses on the streamlining of
the mathematical argument so as to increase our powers of reasoning,
in particular, by the use of formal techniques''.  During the course
he would ask students to write up proofs of the elementary
mathematical problems he discussed during the class and next time he
returned them with detailed comments such as ``Many sins of omissions''.

He was also highly original in his way of assessing people's capacity
for a job. When Vladimir Lifschitz came to Austin in 1990 for a job
interview, Dijkstra gave him a puzzle. Vladimir solved it and
has been working in Austin since then.

Dijkstra's scientific life in Austin was very different from a typical
computer science researcher. To my knowledge he never submitted a
research grant proposal, did not participate in any conference program
committee, and did not attend a conference unless as an invited
speaker.  He read scientific articles mostly by recommendation and
preferred to rely on direct communication with a small group of
colleagues which included some of the most famous computer scientists.
In fact, with many colleagues and friends he maintained a letter
correspondence that occasionally would span a couple of decades.  He
started to use email just a couple of years ago, relying before then
on fax and handwritten letters as means of communication.

\section*{Lifestyle}

As a scientist Dijkstra was a model of honesty and integrity.  Most of
his publications were written by him alone. The few publications that
he wrote jointly with his colleagues bear the unmistakable trait of
his writing style.  He never had a secretary and took care of all his
correspondence alone.  He never sought funds in the form of grants or
consulting and never lent his name to the initiatives to which he
would not contribute in a substantial way.  When colleagues prepared a
festschrift for his sixtieth birthday, published by Springer-Verlag,
he took the trouble to thank each of the sixty one contributors
separately, in a hand-written letter.

His supreme self-confidence went together with a remarkably modest
life style, to the point of being spartan. His and his wife's house in
Nuenen is simple, small and unassuming.  He did not own a TV, a VCR,
or a mobile telephone and did not go to the movies.  In contrast, he
played the piano remarkably well and, while in Austin, liked to go to
the concerts. He also liked to tackle difficult crossword puzzles in
Dutch and would not hesitate to send his solutions to the newspaper.

\section*{Legacy}

Dijkstra's immense intellectual courage and audacity, and deep, yet
strikingly simple and elegant, ideas changed the course of computer
science.  His integrity as a scientist and as a person in private life
cannot be matched.  His views on science in general and on research in
particular were of remarkable depth and originality.  As J Strother
Moore, the chairman of the Computing Science Department at Austin,
said during Dijkstra's funeral: ``He was like a man with a light
in the darkness. He illuminated virtually every issue he discussed.''
In short, he was a genius.  In computer science we are all Dijkstra's
children.

\section*{Acknowledgements}
Ria Dijkstra made us aware that several concepts were introduced by
Dijkstra considerably earlier. Wim Feijen, David Gries 
and Paul
Vit\'{a}nyi offered critical comments on the initial version and
supplied useful corrections and additional information.  Leslie
Lamport provided helpful historical comments.
\end{document}